\documentclass[10pt, oneside,reqno]{amsart}  	% use "amsart" instead of "article" for AMSLaTeX format
\usepackage{geometry}                		% See geometry.pdf to learn the layout options. There are lots.
\usepackage{graphicx}				% Use pdf, png, jpg, or epsÂ§ with pdflatex; use eps in DVI mode
\usepackage{amssymb}
\usepackage[compat=1.1.0]{tikz-feynman} %%% to be used w/ LuaLaTeX %%% 
\usepackage[foot]{amsaddr}
\usepackage{orcidlink}
\usepackage{multicol}
\usepackage{graphicx}  
\usepackage{bbold} 
\usepackage{mathtools}
\usepackage[OT1]{fontenc}
\usepackage{bbding}
\usepackage{amsmath}
\numberwithin{equation}{section} %equation numbering 
\usepackage{amsfonts} 
\usepackage{amssymb}
\usepackage{mathrsfs}
\usepackage{amsbsy}
\usepackage{amsfonts}
\usepackage{physics}
\usepackage{bbold}
\usepackage{slashed}  
\usepackage{wrapfig}
\usepackage{cite}
\usepackage[normalem]{ulem}
\usepackage{hyperref}
 \hypersetup{colorlinks, citecolor=forestgreen, linkcolor=forestgreen, urlcolor=forestgreen}
\definecolor{oucrimsonred}{rgb}{0.6, 0.0, 0.0}
\definecolor{DarkGray}{gray}{0.4}
\definecolor{forestgreen}{rgb}{0.13,0.35,0.13}
\definecolor{ocre}{HTML}{F16723}

\def\eq#1{{Eq.~(\ref{#1})}}
\def\eqs#1#2{{Eqs.~(\ref{#1})--(\ref{#2})}}

\newcommand{\di}{\mbox{d}}

\def\di{\mbox{d}}

\def\}tap{\ \raisebox{-.4ex}{\rlap{$\sim$}} \raisebox{.4ex}{$>$}\ }

\newcommand{\be}{\begin{equation}}
\newcommand{\ee}{\end{equation}}
\newcommand{\bea}{\begin{eqnarray}}
\newcommand{\eea}{\end{eqnarray}}
 
\newcommand{\com}[1]{}
\newcommand{\red}[1]{#1}
\newcommand{\blue}[1]{#1}

\begin{document}

\title{ 
About \red{witnessing} Bell non-locality at colliders}

\author{M. Fabbrichesi$^{\dagger}\,^{\orcidlink{0000-0003-1937-385}}$, R. Floreanini$^{\dagger}\,^{\orcidlink{0000-0002-0424-2707}}$}
\address{$^{\dagger}$INFN, Sezione di Trieste, Via Valerio 2, I-34127 Trieste, Italy}

\author{L. Marzola$^{\ddagger}\,^{\orcidlink{0000-0003-2045-1100}}$}
\address{$^{\ddagger}$National Institute of Chemical Physics and Biophysics, R\"avala pst.10 10143 Tallinn, Estonia, and Institute of Computer Science, University of Tartu, Narva mnt 18, 51009 Tartu, Estonia}
%,  , 
%}

\begin{abstract}%\color{forestgreen}
\red{High-energy colliders enable the testing of quantum mechanics at its most fundamental level, in the presence of strong and electroweak interactions, with systems that consist  of qubits (fermions) and qutrits (massive spin-1 bosons). Quantum state tomography at colliders enables the witnessing of entanglement and Bell non-locality, two defining characteristics of quantum mechanics. We offer a comprehensive explanation of the underlying principles and the methods employed to achieve this remarkable feat.  %The approach also provides a unique perspective on the loopholes exploited by local hidden variable models to circumvent Bell tests.  
} 
\end{abstract}

\maketitle

\section{Introduction}

One of the most unique characteristics of a quantum system is the possibility of \textbf{entanglement}---which refers to the presence of quantum correlations among observables that defy classical physics explanations~\cite{Horodecki:2009zz,Benatti:2010,Nielsen:2012yss,bruss2019quantum}.  Entanglement often clashes with common sense, which is rooted in local realism. Consequently, classical models encompassing standard quantum mechanics have been proposed to explain phenomena in terms of underlying hidden variables. These variables can be deterministic or stochastic, but to replicate the probabilistic nature of quantum mechanics, only their average over a specific distribution influences the outcome of measurements in both the cases.

Investigations into these hypothetical variables sparked lively debates, often delving into intricate philosophical considerations. A significant outcome of these discussions---after von Neumann's rather abrupt dismissal~\cite{vonneumann} was revealed to be premature~\cite{Hermann:1935,Bell:1964fg}---was that hidden variables cannot be indifferent to the experimental setup through which they are being measured. This contextuality requirement arises from a theorem~\cite{Gleason:1957,Bell:1964fg,Kochen:1968zz} that severely limits the possible models of hidden variables. The theorem holds for Hilbert spaces of dimension greater than 2. For a two-dimensional Hilbert space, there exists a straightforward non-contextual model for the spin of a particle~\cite{Bell:1964fg}. Quantum contextuality has been experimentally verified at low energies using photons~\cite{PhysRevLett.84.5457,PhysRevLett.90.250401,PhysRevLett.103.160405,Lapkiewicz:2011rpt,PhysRevLett.109.150401,PhysRevX.3.011012,Ahrens2013}, neutrons~\cite{nature425,PhysRevLett.103.040403}, ions~\cite{Kirchmair2009,PhysRevLett.110.070401,PhysRevLett.120.180401,doi:10.1126/sciadv.abk1660}, molecular nuclear spins~\cite{PhysRevLett.104.160501}, superconducting systems~\cite{Jerger2016}, nuclear spins~\cite{PhysRevLett.123.050401}, and more recently, also at high energies~\cite{Fabbrichesi:2025ifv,Fabbrichesi:2025rsg}.

J. Bell's seminal paper~\cite{Bell:1964} provided a new perspective to the discussion. It demonstrated that a significant portion of these contextual models of hidden variables, those adhering to a locality condition (Bell locality), satisfy specific constraints (\textbf{Bell inequalities}) that can be violated by quantum mechanics~\cite{Brunner:RevModPhys.86.419,scarani2019bell}. This result makes it possible to distinguish between classical models and quantum mechanics through actual experiments~\cite{bell2004speakable,redhead1987incompleteness,bell2002quantum,bertlmann2016quantum,Genovese:2005nw}. Numerous tests of this nature have been conducted using diverse systems at low energies, for instance within atomic and solid-state physics~\cite{Aspect:1982fx,Weihs:1998gy,Tittel:1998ja,Ansmann:2009ims,Hensen:2015ccp,Giustina:2015yza,Storz:2023jjx,PhysRevLett.119.010402}. \red{More recently, entanglement and Bell locality have also been investigated at the higher energies probed in collider experiments (for a comprehensive review, see~\cite{Barr:2024djo}). This unique setting makes it possible  to test quantum mechanics at its most fundamental level, in the presence of both strong and electroweak interactions, as well as with systems that consist not only of qubits, but also of qutrits. 
%seeks to extend the study of Bell locality to high-energy physics by highlighting the relevant provisions. This clarification aims to enhance the understanding of the results obtained.
} 

%Colliders are the only settings where quantum mechanics is tested at its most fundamental level, in the presence of both strong and electroweak interactions. Moreover, particle physics naturally provides examples of systems that consist not only of qubits but also qutrits. 

At particle colliders, it is possible to reconstruct the properties of the particles produced in a collision. For instance, the spin state of the products of a scattering process can be determined by analyzing the angular distributions of the momenta of the related final-state particles, as recorded by a detector. This process, known as \textbf{quantum state tomography}, provides a complete description of the state. %It is important to note that this approach differs from a traditional Bell inequality test, where only joint probabilities along specific directions in space are measured.  For the latter, since only these probabilities are measured, the final state is only partially determined and this lack of knowledge can be exploited to devise \textbf{loopholes} in the test. At colliders, The complete knowledge of the state makes 
\red{Consequently, witnessing Bell non-locality in a bipartite system at colliders is as simple as determining the expectation value of suitable observables on the reconstructed state.}

\blue{To avoid confusion about the meaning of this procedure, it is important to bear in mind the difference between a true \textbf{Bell test} and a \textbf{Bell non-locality witness} that quantum state tomography makes available. The ideal configuration for testing the presence of non-locality is a  \textbf{black box} in which classical inputs can be varied at will (the measurements setting), and classical outputs recorded (the measurement outcomes)~\cite{Eckstein:2021pgm,Altomonte:2024upf}. In this case, measuring a violation of the Bell inequality would support the presence of genuine Bell non-locality in the system. This would then hinder hidden variable models that could recover the experimental result only through possible \textbf{loopholes} specific to the experiment setup. The black box approach has been extensively used in quantum information theory~\cite{Moroder:2013mxj,Scarani:2015vsq,scarani2019bell}, in particular in the context of quantum key distribution protocols~\cite{Acin:2006qom,Ioannou:2021dki}. This approach avoids providing a precise description of the employed equipment, which is often unavailable and would introduce uncertainties due to imperfect measurement settings~\cite{Rosset:2012yzz}. Though most low-energy experiments  are often  device dependent, the black box paradigm can be recovered  after the working of their experimental settings (for example, the polarimeters and the source generating the system of interest) is \textbf{certified} through dedicated experiments.}

The S-matrix approach to quantum field theory was originally inspired by the same black-box principle using the inputs, incoming  particles, and outputs, outgoing particles. This proved fruitful in modeling processes for which the ongoing dynamics was only limitedly understood, as it was the case for the strong interactions. 
%The same approach applies to the creation of the systems used to explore the phenomenology of quantum correlations in collider experiments via quantum state tomography. When analyzing spin correlations, for instance, repeated copies of the selected quantum system are generated in a black-box approach starting from a specific initial state that, in the language of quantum information theory, is trusted. 
At colliders, spin correlations are  reconstructed via a series of tomographic measurements which use, as input, the angular distribution of the momenta of suitable final state particles that reveal the spin of the progenitor particles. The outcome of this procedure is the spin state of the quantum system under investigation, and the presence of properties such as entanglement or Bell non-locality can then be probed through the density matrix formalism with dedicated observables. \blue{Accordingly, the results of a collider experiment will satisfy the conditions for device independence only if the collision dynamics, the detector performance and the reconstruction procedure of quantum correlations are all independently certified. Although this could be---and often it has already been---achieved through a series of independent experiments, the ideal setting of a black-box for performing Bell tests cannot be achieved due to the lack of tunable classical inputs that specify, for instance, the measurement settings. As in collider experiments the settings are essentially chosen by the quantum system under investigation itself, it is necessary to introduce an element of trust on the experimental platform, which makes it impossible to produce certified device-independent results. In the language of quantum information theory, the measurement of Bell non-locality produced at collider experiments via quantum tomography are to be understood as quantum witnesses of non-locality rather than the results of proper Bell tests~\cite{Brunner:RevModPhys.86.419,scarani2019bell} achievable only in with a black-box setting.} 

%Since particle physics is relatively new to this field, we acknowledge the existing jargon and call a Bell non-locality witness what is done at colliders, and with Bell tests what can be done in black-box settings.

\vskip0.5cm
Our discussion follows these steps:
\begin{itemize}
\item[-] We begin with a pedagogical discussion of entanglement in quantum mechanics;
\item[-] This is followed by a discussion of how non-commuting observables are reconstructed from commuting ones;
\item[-] We then provide an overview of quantum tomography---\red{stressing the strengths and weaknesses of the procedure};
\item[-] \red{Finally, we comment on  the actual implementation of hypothetical local hidden models within the framework of a fundamental theory.}
\item[-]  \red{In the Appendix, we explore how the discussion of traditional loopholes is reshaped within the context of collider experiments and quantum tomography.}
\end{itemize}
The paper final section contains  a concise overview of the current \red{possibilities offered by quantum state tomography} at colliders.

\section{Quantum entanglement}

Consider a pair of charged particles that originate from the decay of a neutral state. They fly apart. If, at a later time and in a specific location, you observe a charged particle, you can be certain that somewhere else there will be a particle with the opposite charge. By a measurement  now and here, we gain insights about somewhere else that can be very far away and space-like separated.  However, this does not indicate that the two particles are entangled. Their charge is an intrinsic property that they possess independently of whether we measure it or not. For classical properties, such as all the conserved ones, the knowledge of the distant particle that we instantly acquire by measuring the one close to us simply confirms the existence of classical correlations that are enforced by the neutrality of the progenitor particle.

Instead, genuine quantum correlations originate because of the linearity of the theory, which allows for superpositions of states. Consider the spin of a spin-1/2 particle in a direction determined by the unit vector $\vec n$.  Before the spin is measured, the particle can be in a superposition of the two possible measure outcomes, for instance:
$
|\Psi\rangle = \frac{1}{\sqrt{2}} \, \big( |\!\uparrow_{\vec n} \rangle + |\!\downarrow_{\vec n} \rangle \big) \, ,
$
where $| \!\uparrow_{\vec n} \rangle$ and $| \downarrow_{\vec n} \rangle$ indicate states with the spin aligned or anti-aligned with respect to the direction indicated by $\vec n$, respectively. This means that the component of the spin along a given direction is not an intrinsic property of the fermion. 

A paradigmatic case in which entanglement can be easily studied is precisely that of a bipartite system formed by two spin-1/2 particles, yielding a joint spin state with quantum correlations that can be analyzed. For instance, take a the two-particle state
\be
|\Psi\rangle=\dfrac{1}{\sqrt{2}} \Big(  |  \!\uparrow_{\vec n}\rangle | \!\downarrow_{\vec n} \rangle 
+  | \!\downarrow_{\vec n}\rangle |  \!\uparrow_{\vec n}\rangle \Big) 
\label{superposition}\ ,
\ee 
and repeat the test described before for the charged particles: neither of the two particles carries a definite value for its spin
along the vector $\vec n$, but as soon as one of the two particle is subject to a spin measurement---the result of which is completely random---we learn that the spin of the other lies on the opposite direction.

By taking spin measurements along different directions, the state is not defined by the conservation of the angular momentum and one detects correlations not amenable to a classical explanation. In fact, perfect knowledge of the joint spin state of the particles,
given in (\ref{superposition}), results in a total indetermination of the state of each of the constituent particles. Since this would occur even if the two particles were separated at an arbitrarily large distance at the time of the measurement, one speaks of quantum non-locality.

This brief discussion highlights the central aspect of the quantum mechanical description: particles generally lack definite properties until they are measured. It is not that we lack information about them; rather, it is the peculiar nature of the superposition principle that enables quantum systems to exist in states incompatible with fixed values of these properties before measurements. This starkly contrasts with \textbf{local realism}, which posits the existence of hidden variables to define every property of a particle, representing an element of reality that is locally defined at all times~\cite{Einstein:1935rr}.  
  
%%%%%%%%%%%%

\section{Bell non-locality}

Let us revisit the bipartite system consisting of two spin-$1/2$ particles, where quantum correlations among spin observables can be analyzed and entanglement can be easily studied. In the center-of-mass frame of the bipartite system, the two spin-1/2 particles are moving apart in opposite directions. Upon boosting to the rest frames of each particle, one can then measure the projections of the spin along chosen spatial directions. From these data, one can extract, for instance, the probability $\mathcal{P}(\uparrow_{\vec n}; -)$ of finding the spin of one particle in the state $\ket{\uparrow_{\vec n}}$, aligned to the unit vector $\vec n$, regardless of the spin of the companion particle. Experimentally, this probability can be obtained by looking at the ratio of partial $(\uparrow_{\vec n}; -)$ event counts over the total number of particle pairs. Similarly, one can also consider double probabilities, such as $\mathcal{P}(\uparrow_{\vec n}; \downarrow_{\vec m})$, which represent the probabilities of finding the projection of the spin of one particle aligned with the unit vector $\vec n$ and the companion with spin anti-aligned to a different unit vector $\vec m$. This is precisely what has been measured in all Bell tests conducted so far at low energies using photons, atoms, and quantum dots as qubits.

Let us assume now that the state of the bipartite system can be described by a set of variables that we globally denote with $\lambda$, assumed for simplicity to take continuous values. This description of the system
is also assumed to provide its best characterization.
In particular, the variables $\lambda$ can set the probability $p_\lambda(\uparrow_{\vec n}; -)$
of finding a particle in the state $\ket{\uparrow_{\vec n}}$ regardless of the spin state of the companion particle. Similarly, these variables can set the probability $p_\lambda(\uparrow_{\vec n};\downarrow_{\vec m})$ of finding the
first particle spin in the state $\ket{\uparrow_{\vec n}}$ and the companion particle spin in the state $\ket{\downarrow_{\vec m}}$.

The corresponding average probabilities $\mathcal{P}(\uparrow_{\vec n}; -)$ and $\mathcal{P}(\uparrow_{\vec n}; \downarrow_{\vec m})$,
the only ones accessible to experiments, are  then obtained as
\be
\mathcal{P}(\uparrow_{\vec n}; -)=\int d\lambda\ \eta(\lambda)\ p_\lambda(\uparrow_{\vec n}; -)\ ,\quad
\text{and} \quad \mathcal{P}(\uparrow_{\vec n}; \downarrow_{\vec m})=\int d\lambda\ \eta(\lambda)\
p_\lambda(\uparrow_{\vec n};\downarrow_{\vec m})\ ,
\label{probabilities}
\ee
with the probability density distribution $\eta(\lambda)$, normalized as $\int d\lambda\ \eta(\lambda)=1$, which characterizes the ensemble of bipartite systems observed in the experiment. It is important to stress that this description is very general and can be made to agree with the prediction of quantum mechanics.

We shall now focus on the situation, common in classical probability theory,
for which the two probabilities $p_\lambda(\uparrow_{\vec n}; -)$
and $p_\lambda(-; \downarrow_{\vec m})$ are assumed to be independent.
In other terms, we shall consider a description of the two spin-1/2 system for which:
\begin{equation}
\boxed{p_\lambda(\uparrow_{\vec n};\downarrow_{\vec m})=
p_\lambda(\uparrow_{\vec n}; -)\ p_\lambda(-; \downarrow_{\vec m})\ .}
\label{locality}
\end{equation}
This assumption is known as \textbf{Bell locality condition}~\cite{bell2004speakable,redhead1987incompleteness,bell2002quantum,bertlmann2016quantum}.
Measurable consequences of this assumption can be easily derived~\cite{Clauser:1974tg}.

First of all, note that for any set of four non-negative numbers $x_1$, $x_2$, $x_3$ and $x_4$,
less or equal than 1, the following inequality holds:
\begin{equation}
x_1 x_2 - x_1 x_4+ x_3 x_2 +x_3 x_4 \leq x_3 +x_2\ .
\label{number-inequality}
\end{equation}
Replacing $x_i$ with suitable probabilities $p_\lambda$ and integrating over $\lambda$ with the weight
$\eta(\lambda)$, from (\ref{number-inequality}) one can deduce the following relation:
\begin{equation}
\mathcal{P}(\uparrow_{{\vec n}_1}; \uparrow_{{\vec n}_2}) - \mathcal{P}(\uparrow_{{\vec n}_1}; \uparrow_{{\vec n}_4})
+\mathcal{P}(\uparrow_{{\vec n}_3}; \uparrow_{{\vec n}_2}) + \mathcal{P}(\uparrow_{{\vec n}_3}; \uparrow_{{\vec n}_4})
\leq\mathcal{P}(\uparrow_{{\vec n}_3}; -) + \mathcal{P}(-; \uparrow_{{\vec n}_2})\ ,
\label{Bell-inequality}
\end{equation}
where ${\vec n}_1$, ${\vec n}_2$, ${\vec n}_3$ and ${\vec n}_4$
are four three-dimensional unit vectors determining four arbitrary spatial directions
along which the spins of the two particles can be measured. This condition involving double probabilities $\mathcal{P}$ is a particular instance of a Bell inequality.
It is worth stressing that due to the great generality used in its derivation, the relation (\ref{Bell-inequality}) must be satisfied by all local deterministic or stochastic alternatives to quantum mechanics---that is,
models in which the condition (\ref{locality}) is assumed to hold.

\com{Let us now discuss what ordinary quantum mechanics predicts for the combinations of probabilities appearing in (\ref{Bell-inequality}).
A bipartite quantum system made of two qubits, spin-1/2 particles in this case,
is described in terms of the $4$-dimensional Hilbert space 
$\mathcal{H}_4=\mathcal{H}_2\otimes\mathcal{H}_2\equiv\mathbb{C}^4$ given by the tensor product of the two \hbox{$2$-dimensional}
Hilbert spaces, $\mathcal{H}_2\equiv\mathbb{C}^2$, each containing the possible states of a single qubit. Any observable $\mathcal{O}$ of the full system can then be expressed in a tensor product form,
$\mathcal{O}=\mathcal{O}_1\otimes\mathcal{O}_2$, where $\mathcal{O}_1$, $\mathcal{O}_2$ are
single-spin observables such as, for instance, spin projections along different spatial directions.}

The spin state of the bipartite system can be  encoded in a density operator $\rho$ represented by a $4\times 4$ matrix
that must be semi-positive and of unit trace in order to assure the correct interpretation of its eigenvalues $\rho_i$, $i=1,2,3$, as probabilities: $\rho_i\geq0$, ${\rm Tr}[\rho]=\sum_i \rho_i =1$. It can be expanded as
\begin{equation}
\rho=\frac{1}{4}\Big[{\bf 1}\otimes{\bf 1} + \sum_i B_i^{(+)} (\sigma_i\otimes{\bf 1})
+ \sum_j B_j^{(-)}({\bf 1}\otimes \sigma_j) + \sum_{ij} C_{ij} (\sigma_i\otimes\sigma_j) \Big]\ ,
\label{rho}
\end{equation}
where $\sigma_i$ are the Pauli matrices, $\bf 1$ is the unit $2\times 2$ matrix, and where the sums involving the indices $i$ and $j$ run over the labels $1$, $2$ and $3$, representing any orthonormal spatial reference frame. The real coefficients $B_i^{(+)}={\rm Tr}[\rho\, (\sigma_i\otimes {\bf 1})]$ and
$B_j^{(-)}={\rm Tr}[\rho\, ({\bf 1}\otimes\sigma_j)]$ give the spin averages (\textit{i.e.} the polarizations) of the corresponding particles, while the real matrix $C_{ij}={\rm Tr}[\rho\, (\sigma_i\otimes\sigma_j)]$ contains the spin correlations. Note that while the density matrix in (\ref{rho}) is normalized to ${\rm Tr}[\rho]=1$, extra constraints on $B_i^{(+)}$, $B_j^{(-)}$ and $C_{ij}$ are needed to guarantee non-negative eigenvalues. These extra conditions are in general non-trivial, requiring 
all principal minors of $\rho$ to be non-negative. Knowledge of $\rho$ allows to compute the average of any two-spin observable $\mathcal{O}$
as $\langle\mathcal{O}\rangle={\rm Tr}[\rho\, \mathcal{O}]$, including
the probabilities appearing in (\ref{Bell-inequality}).

 The use of a density matrix does not entail that quantum mechanics is necessarily being used, but simply that the utilized quantities  have no fixed values and follow instead  a probabilistic distribution---the same as in statistical mechanics or classic optics, in which density matrices are often used. Also the Pauli matrices appearing in \eq{rho}---which seems to bring in quantum mechanics---are just a convenient basis in the $2\times 2$ space of the polarizations.

The probability for a single spin to be in a state $\ket{\uparrow_{\vec n}}$ or $\ket{\downarrow_{\vec n}}$ along the direction $\vec n$ can be obtained by computing the average over the spin state of the following two
projection operators, respectively:
\begin{equation}
P_{\uparrow_{\vec n}}=\frac{1}{2}\big({\bf 1} + {\vec n}\cdot{\vec \sigma}\big)\ ,\qquad
P_{\downarrow_{\vec n}}=\frac{1}{2}\big({\bf 1} - {\vec n}\cdot{\vec \sigma}\big)\ ,\quad
\text{and} \quad {\vec n}\cdot{\vec \sigma}\equiv\sum_{i=1}^3 n_i\, \sigma_i\ .
\label{spin-projectors}
\end{equation}
Then, in the case of the two-spin system, one similarly has:
\begin{equation}
\mathcal{P}(\uparrow_{\vec n}; -) = {\rm Tr}\Big[\rho\, \Big(P_{\uparrow_{\vec n}}\otimes{\bf 1}\Big)\Big]\ ,\qquad
\mathcal{P}(-; \uparrow_{\vec m}) = {\rm Tr}\Big[\rho\, \Big({\bf 1}\otimes P_{\uparrow_{\vec m}}\Big)\Big]\ ,
\label{single-probabilities}
\end{equation}
as we here disregard the spin state of the companion particle. Analogously, in the case of the double probability
$\mathcal{P}(\uparrow_{\vec n}; \uparrow_{\vec m})$, one has:
\begin{equation}
\nonumber
\mathcal{P}(\uparrow_{\vec n}; \uparrow_{\vec m})=\
{\rm Tr}\Big[\rho\, \Big(P_{\uparrow_{\vec n}}\otimes P_{\uparrow_{\vec m}}\Big)\Big]\ .
\label{double-probabilities}
\end{equation}
Inserting these results into the inequality (\ref{Bell-inequality}) leads to the following algebraic relation:
\begin{equation}
{\vec n}_1\cdot C \cdot \big({\vec n}_2 - {\vec n}_4 \big) +
{\vec n}_3\cdot C \cdot \big({\vec n}_2 + {\vec n}_4 \big)\leq 2\ .
\label{algebraic-relation}
\end{equation}
involving only the correlation matrix $C$. Combining this condition with the analogous one obtained by replacing $\uparrow_{{\vec n}_1}$ with $\downarrow_{{\vec n}_1}$
and $\uparrow_{{\vec n}_3}$ with $\downarrow_{{\vec n}_3}$ in \eq{Bell-inequality}, we finally obtain the following inequality:
\begin{equation}
\Big|{\vec n}_1\cdot C \cdot \big({\vec n}_2 - {\vec n}_4 \big) +
{\vec n}_3\cdot C \cdot \big({\vec n}_2 + {\vec n}_4 \big)\Big|\leq 2\, .
\label{algebraic-inequality}
\end{equation}

In order to put under test the Clauser-Horne-Shymony-Holt (CSHC) inequality~\cite{Clauser:1969ny} in \eq{algebraic-inequality}\footnote{\red{This inequality differs from the original Bell inquality by being about correlations rather than probabilities.}}, one needs to experimentally determine the matrix $C$ and then suitably choose four spatial directions ${\vec n}_1$, ${\vec n}_2$, ${\vec n}_3$ and ${\vec n}_4$
that maximize the left-hand side of \eq{algebraic-inequality}. Fortunately, this maximization process can be easily performed analytically, for a generic spin correlation matrix. This result has been demonstrated in~\cite{Horodecki:1995340}. Indeed, consider the matrix $C$ and its transpose $C^T$ and form the symmetric, positive, $3\times 3$ matrix
$M= C^T C$; its three eigenvalues $m_1$, $m_2$, $m_3$ can be ordered in decreasing order: $m_1\geq m_2\geq m_3$. Then, the following result holds:
the two-spin state $\rho$ in \eq{rho} violates the inequality (\ref{algebraic-inequality}), or equivalently (\ref{Bell-inequality}),
if and only if the sum of the two largest eigenvalues of $M$ is strictly larger than 1, that is
\be
\boxed{\mathfrak{m}_{12}\equiv m_1 + m_2 >1\,.}
\label{Horodecki-crit}
\ee
In other terms, given a spin correlation matrix $C$ of the state $\rho$ that satisfies $\mathfrak{m}_{12} >1$, then there is at least a choice of the four  vectors ${\vec n}_1$, ${\vec n}_2$, ${\vec n}_3$, ${\vec n}_4$
for which the left-hand side of \eq{algebraic-inequality} is larger than 2.

\blue{As mentioned in the introduction, it is important to stress how the above procedure differs from the conventional approach commonly employed in testing Bell inequalities, particularly within the field of quantum optics. In standard Bell tests, the probabilities that enter the inequality (\ref{Bell-inequality}) are experimentally determined by measuring the frequencies of events for each choice of the directions ${\vec n}_i$. In the above approach, instead, we only rely on the experimental determination of the spin correlation matrix $C$ provided by quantum state tomography. As no experimental setting can be arbitrarily varied, it is necessary to trust that the angular distributions used to reconstruct the correlation matrix are produced by a fair source.}
\blue{To the extent that in quantum tomography both outcomes and the settings are determined together and not independently from each other, this approach provides, in the language of the quantum information theory, a witness of Bell non-locality and not a Bell test. }

\section{Quantum information observables in the $S$-matrix framework}

A superficial comparison with the low-energy tests of quantum mechanics based on polarized photons and polarimeters might suggest that, in order to explore quantum entanglement and Bell \red{non-locality} violation at colliders, one must necessarily use polarized beams and polarized targets. Yet this is not the way high-energy physics has developed and there are essentially no experiments using such beams and targets. %
%%%%%%%%%%%%%%%%%%
%\begin{wrapfigure}{L}{0.5\textwidth}
 \begin{figure}[h!]
\begin{center} 
\com{\vskip-0.25cm
\feynmandiagram [large, horizontal=a to b] {
i1 [particle=\(e^{-}_L\)] -- [fermion] a -- [fermion] i2  [particle=\(e^{+}_R\)],
a -- [photon, edge label=\(\gamma\)] b,
f1 [particle=\(\tau^{+}_L\; \mbox{or} \; \tau^{+}_R\)] -- [fermion] b  -- [fermion] f2 [particle=\(\tau^{-}_R\; \mbox{or} \; \tau^{-}_L\)],
};}
   \includegraphics[width=0.5\textwidth]{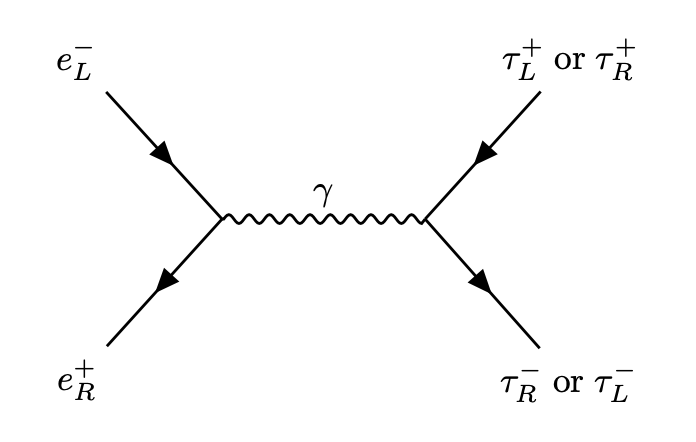}
   \caption{\small \label{fig:tau} Feynman diagram for the the production of a pair of $\tau$ leptons from electron-positron annihilation.} 
\end{center}
\end{figure}
%\end{wrapfigure}
%%%%%%%%%%%%%%%%%%%%%%
Almost all experiments are, instead, based on (mostly) unpolarized beams and detectors that do not measure directly polarizations. Is this situation a show stopper\footnote{See, for example, \cite{Abel:1992kz}.  \red{ The argument in this paper hinges on the existence of a local hidden variable model---introduced by Kasday\cite{kasday}---in which the hidden variables are directly related to the momenta and their distribution given by the Standard Model differential cross section. As explained in this work, this implies that the model is an instance of the class of super-deterministic models which are discussed  in Section 5.}} for testing quantum mechanics---be it Bell non-locality or entanglement---at colliders? Not at all---as the rich physics program on polarization effects at colliders shows. Polarizations are reconstructed from the angular distributions of particles in the final state.  As explained in detail below, the  momentum angular distribution is used to estimate correlations, thereby bringing us from the cross section to a density matrix. It is in the space in which this matrix is defined that entanglement and Bell \red{non-locality} must be studied---not in the cross section. This is a  powerful way of studying polarizations and it allows to measure also the polarization of particles that are not directly seen in the detector because they decay sufficiently fast. \blue{Studies of entanglement and Bell non-locality in isospin space, as reported in~\cite{Fabbrichesi:2025rqa}, further support the notion that quantum mechanics can indeed be tested using quantum state tomography. Notably, this approach eliminates the need for momentum measurements, and the final state is entirely determined by the values of the branching rates in the various channels.}

Consider the simplest process at colliders, namely the production of a pair of $\tau$ leptons from electron-positron annihilation (see Fig.~\ref{fig:tau}). We take $\tau$ leptons because their polarizations can be reconstructed by analyzing the angular distributions of suitable decay products. The final state, in general, is given by a superposition of bipartite $\tau$ lepton states that span the four possible $R$, $L$-helicity combinations:
%\vskip0.3cm
\be
|\Psi\rangle=\zeta_1 | \tau^-_L\rangle | \tau^+_L \rangle + \zeta_2 | \tau^-_R\rangle | \tau^+_L \rangle
 + \zeta_3 | \tau^-_L\rangle | \tau^+_R \rangle+ \zeta_4 | \tau^-_R\rangle | \tau^+_R \rangle\,,
\quad \text{with} \quad \sum_i |\zeta_i|^2 =1\, .
\ee
Denoting with $\Theta$ the $\tau^+$ scattering angle in the centre-of-mass frame of reference, a direct computation within the Standard Model gives
\be
\boxed{
|\Psi\rangle = \frac{1}{\sqrt{2}}\, \frac{ 1 + \cos \Theta}{\sqrt{1+\cos^2 \Theta}} | \, \tau^-_R\rangle | \tau^+_L \rangle + 
 \frac{1}{\sqrt{2}}\ \frac{ 1 - \cos \Theta}{\sqrt{1+\cos^2 \Theta}} | \,  \tau^-_L\rangle | \tau^+_R \rangle\ ,}
\label{tau-state}
\ee
because of the vector-like photon coupling. 

The angular distribution of the charged pions produced in the decays $\tau^\pm \to \nu_\tau \pi^\pm$ can be used to reconstruct the polarizations and spin correlation of the tau pairs. The connection is provided by the conservation of the angular momentum in the rest frame of the decaying lepton.

For a single $\tau$ lepton, the polarization density matrix is given by
\be
\rho [s_i] = \begin{pmatrix}
\frac{1}{2} + \langle s_z \rangle & \langle s_x+i s_y \rangle\\
 \langle s_x- i s_y \rangle & \frac{1}{2} - \langle s_z \rangle 
\end{pmatrix}\ , \label{pol1}
\ee
in which $\expval{s_i}$ are the average polarization components in  a given basis.

As already discussed, the use of the density matrix does not imply that quantum mechanics is being used. For example, in optics
the same  density matrix is often used to describe a  beam of light as
\be
\rho_{\text{light}} = \dfrac{1}{2} \begin{pmatrix}
1+ \xi_3  & \xi_1 - i\, \xi_2\\
\xi_1 + i \,\xi_2&  1- \xi_3
\end{pmatrix}\ ,
\ee
in terms of the Stokes parameters $\xi_i$---each of them characterizing linear  and circular polarizations along given directions.

 The expectation values of the polarization components in \eq{pol1}  are reconstructed from the events that give rise to the differential cross section
\be
\frac{1}{\sigma} \frac{\di \sigma}{\di \cos \theta_i^\pm} = \frac{1}{2}  \Big( 1 +  \langle s_i \rangle \, \alpha_i \,\cos \theta_i \Big)\ , \label{polarimeter}
\ee
with $\theta_i$ the angles described by the momenta of the charged pions in the decay of the $\tau$ lepton.

  This reconstruction, where the decaying particle acts as its own polarimeter, relies on the independent experimental determination of the polarimetric vector. \blue{It is commonly performed for hyperons and other heavy baryons in charmonium decays~\cite{BESIII:2020fqg,BESIII:2022qax,BESIII:2025vsr}. This reconstruction may not be available from the data alone for particles that are not produced in a definite state of polarization.}  This is the case of the $\tau$ particle at present colliders, for which the coefficient $\alpha_i$ in \eq{polarimeter} turns out to be $\alpha_i=1$ for the single-pion $\tau$ decay, solely due to the conservation of angular momentum in the decay process.

 %This determination is mirrored in low-energy experiments with photons by the measurement of the properties of the polarimeter used in the experimental setup. Both experiment types require an external certification from experiments that are independent of those testing Bell non-locality. 

 \begin{figure}[h!]
  \begin{center}
    \includegraphics[width=0.43\textwidth]{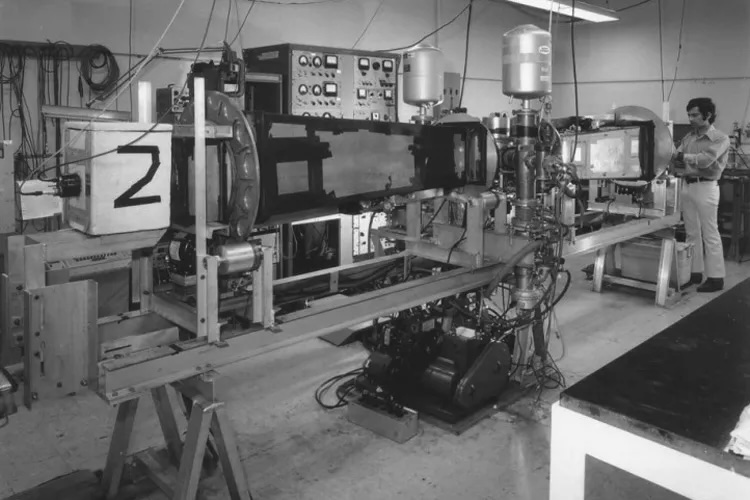}
      \includegraphics[width=0.51\textwidth]{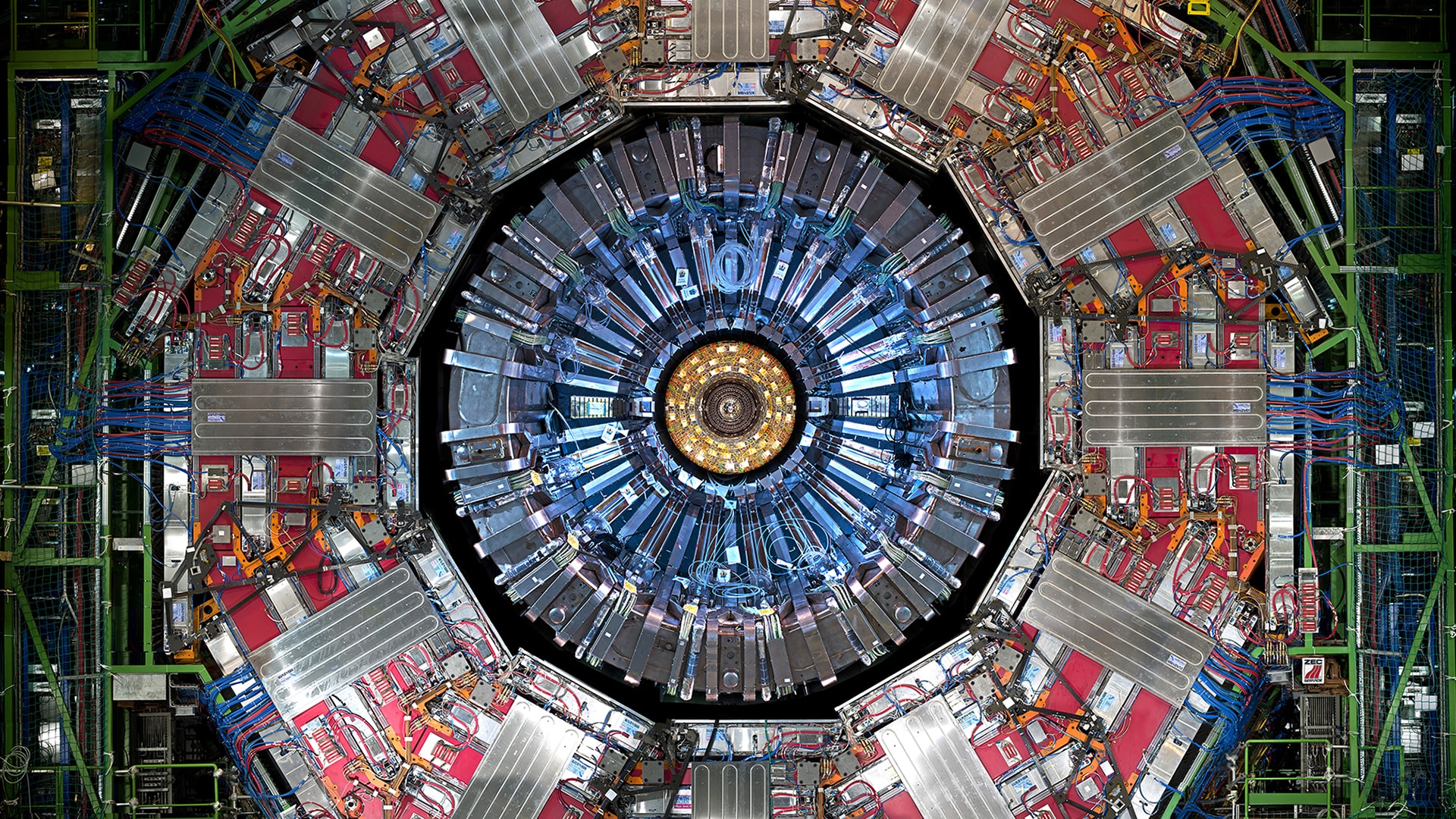}
  \end{center}
 \caption{\small \label{fig:exp} Clauser's experimental setup (Stuart Freedman, Berkeley  [CC BY-NC 4.0]) and the CMS detector at the LHC (CERN  [CC BY-NC 4.0]). Neither is as simple as a falling apple and the device-independence of potential Bell test has to be certified.}
\end{figure}
%%%%%%%%%%%%%%%%%%%%%%%%%%%%%%%%%

All experiments---be them  at low or high energy  (see Fig.~\ref{fig:exp})---must necessarily rely on working of the polarimeter utilized in the measurements. It is the polarimeter that transforms a measured data (light or darkness in the case of photons, angular distribution of momenta in the case of the decay of a particle) into a polarization measurement. In the case of the atomic-physics tests, this equipment consists of the photon polarimeter, while  in the case of the high-energy tests, in the polarimetric vector and spin-analyzing power of a particle emitted in a decay. \blue{In both cases, the actual operation of the polarimeter has to be certified in independent experiments apt to provide the characteristic properties of the photon polarimeter or of the polarimetric vector.\footnote{It is worth pointing out that in an analytic computation, or a Monte Carlo simulation, the operation of the polarimeter can be determined directly from the specific features of the model used, rather than being derived from some other data. For instance, in the context of tests at colliders, the polarimetric vector in the fully leptonic decay of the top quark is found to be equal to one due to the chiral coupling between elementary fermions and gauge bosons in the Standard Model.} It is only after this step that the experimentally collected data---be them in the form of frequencies, probabilities or correlations---can be processed to create device-independent results. %\blue{For the case of high-energy physics, we remark that the extraction of the data does not in general assume quantum mechanics, and where it may depend from quantum mechanics (the decay and transfer of the polarization to the decay final state) can be analyzed in experiments on single particle decay which are independent from those on the bipartite system. The same is done in the traditional settings in which the polarimeter (whose working is base on Quantum Mechanics) is trusted because tested in independent experiments.  
} In regard of this, experiments always compare models against the data. The data represent the ``real world,'' and when presented with a model, we assess how well the data aligns with it. However, there is no direct way to test the ``real world.'' Consequently, the Bell test simply compares the factorization of probabilities (Bell locality) against the data. Its results can be used to rule out local theories (if the inequality is violated) and even quantum mechanics (if the violation exceeds a critical value known as Tsirelson's bound~\cite{Cirelson:1980ry}). The same principle applies to measurements of entanglement.

For the final state with two $\tau$ leptons, the differential cross section is given by
\be
\frac{1}{\sigma} \frac{\di \sigma}{\di \cos \theta^+_i \di \cos\theta^-_j}= \frac{1}{2}  \Big( 1 + \alpha_i B^+_i \cos \theta^+_i+ \alpha_i B^-_i \cos \theta^-_i +C_{ij}  \alpha_i \alpha_j \cos \theta^+_i  \cos \theta^-_j \Big)\ ,
\ee
where the angles $\theta_i^\pm$ give the spatial orientation of the charged pion momenta in the rest frames of the progenitor $\tau$ leptons with respect to some basis with components $i$ and $j$. The coefficients $B^\pm_i$ and $C_{ij}$ can be included
into the polarization density matrix $\rho [C_{ij}\, , B_i]$, as in (\ref{rho}). The logical steps leading to the determination of the average
of any system observable $\mathcal{O}$ is then as follows:
\be
\boxed{\biggl  \langle 
\alpha_i \alpha_j \cos \theta_i \cos \theta_j \biggr \rangle \quad  \Rightarrow
\quad C_{ij}\, , B_i ^{\pm} \quad  \Rightarrow \quad
\rho [C_{ij}\, , B_i^{\pm}]   \quad  \Rightarrow \quad \langle \mathcal{O} \rangle = {\rm Tr}[\rho\, \mathcal{O}]\ .}
\ee
The measurement provides the angular distributions of interest, in the first term above, averaged over the overall distribution of the events.  From there, one extracts polarization and correlation coefficients, $B_i$, $C_{ij}$, that can be organized in the density matrix $\rho$, then used to determine averages of physical quantities.
The polarization state of the two $\tau$ lepton is completely reconstructed: this is done through quantum state tomography, by a likelihood fit of the angular distribution of the decay products.
Once the state is given, its properties can be obtained by means of averages of suitable observables, like the concurrence $ {\mathcal C}[\rho] $ for quantifying its entanglement content~\cite{Bennett:PhysRevA.54.3824,Wootters:PhysRevLett.80.2245,Rungta:PhysRevA.64.042315}, 
or testing the Bell \red{non-locality} (\ref{algebraic-inequality}) through the criterion (\ref{Horodecki-crit}).\footnote{
The density matrix reconstructed by the experiments is a convex combination of the density matrices proper of the involved states, each characterized by a different kinematic configuration. As a specific example, consider again an ensemble of $\tau$ lepton pairs produced in repeated $e^+\, e^-$ collisions, each event being characterized by a different value of the scattering angle. Without imposing any kinematic cut, an experiment would then reconstruct the matrix
\begin{equation}
  \overline\rho = \frac{\int_\Omega \dd\Omega \frac{\dd\sigma}{\dd\Omega} \rho(\Omega)}{\int_\Omega \dd\Omega \frac{\dd\sigma}{\dd\Omega}} = \frac{1}{\sigma}\int_\Omega \dd\Omega \frac{\dd\sigma}{\dd\Omega} \rho(\Omega)
  \label{eq:rhoav}
\end{equation}      
where $\Omega$ is the solid angle defined by the direction of one of the leptons in the center of mass reference frame. Does the above density matrix describe a valid quantum state? The answer is: it depends~\cite{Afik:2022kwm,Cheng:2023qmz,Cheng:2024btk}.
To understand why, notice that the quantum tomography procedure relies on a triad of orthonormal vectors, explicitly used in the summation appearing in  \eq{rho}, to define the orientation in space of the spin vectors of interest. If the same triad is used for all the events, for instance the three directions $\hat{x}$, $\hat{y}$, $\hat{z}$ in the center of mass frame, then the matrix in \eq{eq:rhoav} describes a genuine quantum state. On the contrary, when the triad is picked event by event, as with the $\hat n$, $\hat r$, $\hat k$ tern related to helicity frames, the matrix $\overline \rho$ is an average referred as \textbf{fictitious states}. We remark that the Fano coefficients $\overline{B}^\pm_i$ and $\overline{C}_{ij}$ obtained from $\overline\rho$, for instance, through the decomposition in \eq{rho} are still informative: they correspond to the ensemble averages of the corresponding quantities for the chosen basis. Furthermore, because separability and Bell-locality are maintained in convex combinations, the density matrix in \eq{eq:rhoav} can always be used to highlight the presence of entanglement or Bell \red{non-locality} in some of the states involved in the experimentally-reconstructed average. }  Again, the use of the density matrix in estimating correlations is not proper to quantum mechanics: it is commonly used also in statistical physics, where it gives the distribution probability in phase space.

%%%%%%%%%%%%%%%%%%%%%%
Does this reconstruction provide a non-trivial measurement of the polarizations? It does. The events could give rise to a separable state instead of an entangled one. This is the case if the distribution of probability in $\cos \theta_i \cos\theta_j$ factorizes in the product of the cosines. Do the found values  test quantum mechanics? Yes,  the events  are measured independently of quantum mechanics and could give rise to a state more entangled than allowed by quantum mechanics. This is the case, for example, if some interaction (a reborn Maxwell's demon) introduces a modified  event distribution such that the correlations come to exceed the Tsirelson's bound~\cite{Cirelson:1980ry}  predicted by quantum mechanics. A more explicit example is provided by supersymmetric entangled qubits~\cite{Borsten:2012pp}. Only the correlations $C_{ij}$ enter in the violation of the bound and no property of the density matrix is necessarily implied. As a matter of fact, it is doubtful whether a density matrix can be properly defined when the correlations exceed the Tsirelson's bound. Finally, also mixed states that are entangled but local (Werner's states~\cite{PhysRevA.40.4277}) can and do arise in this approach.

If the reader feels uneasy about using commuting observables, such as momenta, to reconstruct a non-commuting observable  like the spin, consider that \red{observed quantities are always real numbers and, as such, commuting.} For instance, in the Stern-Gerlach experiment, the spin of a neutron is measured by its position (or momentum) on a screen. The commuting variable simply provides the spin value, which is either up or down. It is only through repeated experiments using different directions and comparing the results for the neutron's spin that its non-commutative nature is revealed.

Similarly, at colliders, specific momentum angular distributions are used as  distribution of probabilities to estimate correlations, thereby bringing us from the cross section to the density matrix. It is in the Hilbert space on which this matrix acts that entanglement and Bell \red{non-locality}  must be studied---not in the cross section.

To elaborate this point into an explicit computation, we can observe from equation \eq{tau-state} that as we vary the scattering angle, we can obtain the separable states. 
\be
 | \tau^-_R\rangle | \tau^+_L \rangle \quad \text{or} \quad | \tau^-_L\rangle | \tau^+_R \rangle \label{sep}\ ,
\ee
respectively for $\Theta=0$ and $\Theta=\pi$, enforced by the conservation of angular momentum. The maximally entangled state
\be
|\Psi\rangle=\dfrac{1}{\sqrt{2}} \Big(  | \tau^-_R\rangle | \tau^+_L \rangle+  | \tau^-_L\rangle | \tau^+_R \rangle \Big)\ ,
\label{bell}
\ee 
reminiscent to that of \eq{superposition}, is found instead for $\Theta=\pi/2$. 

The amount of entanglement admitted by the state can be quantified, for instance, by the \textbf{concurrence} ${\mathcal C}[\rho]$ (see \cite{Guhne:2008qic} for a discussion of more general entanglement witnesses). For the $\tau$-lepton bipartite system this is given by
\be
{\mathcal C}[\rho]= \frac{\sin^2 \Theta}{1 + \cos^2\Theta}\, ,
\ee
and, as expected, equals 1---its maximal value---for $\Theta=\pi/2$. 

In the massless limit we consider, the chirality states coincide with the helicity ones, hence the Bell state in \eq{bell}  is completely analogous to that formed by two photons emitted by an atom in a state of zero total angular momentum. The two photons are in the entangled state in \eq{bell}, involving now the two helicity states, $L$ and $R$, of the photons $(1)$ and $(2)$:
\be
|\Psi\rangle=\dfrac{1}{\sqrt{2}} \Big(  | \gamma_R^{(1)}\rangle | \gamma_L ^{(2)}\rangle+  | \gamma_L^{(1)}\rangle | \gamma_R^{(2)} \rangle \Big)\,.\label{ph}
\ee

Notice  that the polarizations of both the massless leptons and the photons are described by the Wigner little group $ISO(2)$, which reduces to the Abelian $U(1)$ after imposing the restriction on finite representations. The non-commuting nature of the polarization observables shows up---for the photons as well as for the $\tau$ leptons---in the four-dimensional space of their eigenstates. In this space the state $\Psi$ can be written as a four-vector
\be
\Psi =\dfrac{1}{\sqrt{2}} \begin{pmatrix} 0\\1\\1\\0\end{pmatrix} \label{state1}
\ee
in the case of \eq{bell} and
\be
\Psi = \begin{pmatrix} 0\\0\\1\\0\end{pmatrix} \quad \text{or} \quad
\Psi = \begin{pmatrix} 0\\1\\0\\0\end{pmatrix} \label{state2}
\ee
in the case of \eq{sep}.\footnote{\blue{This is an essential point: the helicities (or the polarizations), which are real numbers and therefore commuting quantities, only show their non-commuting properties after being organized in the corresponding Hilbert space in the states in \eqs{state1}{state2}. Even if the cross section were to explicitly depend on the spin variable and the detector contained a polarized target, measurements could still rely on the final momentum distribution to reconstruct the direct effect of the spin interactions.}
} Once the state is known, tests of Bell locality or any other property are just a matter of taking expectation values of the appropriated operators. Here we mimic those leading to the traditional setting of a Bell test, by studying the inequality (\ref{Bell-inequality}).

We can explicitly test that a separable state, such as the first of those in \eq{sep}, gives in the just specified basis, for $i,j=1,2,3$,
 \be
  {\mathcal P} (\uparrow_{\vec n_i}; \uparrow_{ \vec n_j}) = \frac{1}{4} ( 1 -  n^z_i +  n_j^z - n_i^z\ n_j^z )\, ,
  \ee
  with 
  \be
  {\mathcal P} (\uparrow_{\vec n_i}; -)= \frac{1-n^z_i}{2} \quad \text{and} \quad   {\mathcal P} (-; \uparrow_{\vec n_j})= \frac{1+n^z_j}{2} \ .
  \ee
  The non-separable state in \eq{bell} gives
  \be
 {\mathcal P} (\uparrow_{\vec n_i}; \uparrow_{\vec n_j}) =  \frac{1}{4} (1 +  n_i^x  n_j^x + n_i^y  n_j^y- n_i^z  n_j^z )\,, 
  \ee
 and  1 for the sum of the single spin probabilities.

  The choice
  \be
  \vec n_1 = \vec z\, , \quad \vec n_2=\frac{-1}{\sqrt{2}} (\vec z + \vec x)\, , \quad \vec n_3=-\vec x\, \quad \vec n_4 = \frac{1}{\sqrt{2}} (\vec z - \vec x)
  \ee
gives
 \be
 \frac{1}{2} \leq 1-\frac{1}{2 \sqrt{2}}\,,
 \ee
 in the first case, satisfying the inequality. In the second case, the same choice leads instead to
\be
\frac{1}{2} + \frac{\sqrt{2}}{2} \nleq 1 
\ee
and the consequent violation of the inequality.
   
In this demonstration, the polarizations of the $\tau$ leptons have been identified with their helicities in the massless limit, to make the analogy with photons clearer. The fact that momenta are commuting variables is irrelevant because it is the non-commuting nature of the spin polarization measurements in the $2\times 2$ space of the spins that enters the Bell inequality.
   
This derivation explicitly shows how Bell non-locality can be \red{probed} at a collider experiment and how the non-separable, Bell state violates the inequality. It follows the traditional protocol of fixing some directions and comparing probabilities. With quantum tomography this is only an afterthought since the reconstruction of the state automatically shows the presence or absence of entanglement and the implied Bell \red{non-locality}---having the state we can perform any test we wish on it to highlight its quantum properties.

\section{\red{Quantum state tomography and hidden variables in collider experiments}}    

 %\blue{When studying spin correlations at collider experiments, the variables that are locally or non-locally entangled are the spins of the particles (or the helicities, for massless states), not their momenta. The latter are used in the reconstruction of spin correlations and to study their properties we must go to their Hilbert space (since discussing whether an observable commutes or does not is only possible in its Hilbert space).The same is true for traditional Bell tests with the photon helicities organized in the  polarization state in \eq{ph}.}
 
\red{The investigation of non-locality performed with quantum state tomography cannot be regarded as a Bell test as it is inherently open to a fundamental loophole. This arises from the lack of control on the angular directions used by the tomographic procedure and  consists in assuming the existence of a hidden variable distribution that contains the outcome of the experiment that is about to be performed. Alternatively, this distribution could be set to match the Standard Model predictions~\cite{kasday}, which are known to agree with experiments very well at the current precision. In either case, for particle physics experiments, this would imply that the hidden variables regulate the angular distribution of the momenta used by the tomographic procedure, in such a way that local hidden variable models could mimic the effects of a genuine non-local behavior. 
Furthermore, the same variables must also convey information about the detector, such as the direction of the magnetic field used to differentiate the charges of particles, due to the contextuality requirement~\cite{PhysRevLett.48.291}.}

\blue{As the collision dynamics has been certified to follow the principles of quantum field theory, all the information about the final momentum distribution and the experimental setup used must be encoded in the initial state formed by the colliding particles. This loophole then represents an instance of \textbf{super-determinism}, a philosophical stance that suggests in this case that hidden variables determine both the behavior (encoded in  probability distributions) of the particles as well as the choices made by experimenters in a traditional Bell test. All Bell tests are inherently susceptible to this loophole~\cite{Brans:1987} because the freedom in choosing the measurement directions---the free will of experimenters, in other words---is one of the assumptions underlying the derivation of the Bell inequality.}

\red{Super-deterministic models, even in their milder version in which the free will is not involved and only probabilistic distributions are to be determined beforehand, cannot be ruled out with scientific methods, neither at collider experiments nor with more traditional setups using atoms or solid-state devices. Yet, they appear to be incompatible with our  understanding of the physical world as they necessitate the absolute determination of the initial conditions~\cite{potter2025} which, in turn, requires specifying all the infinitely many digits comprising the real numbers that describe these conditions~\cite{PhysRevA.100.062107}. That a complete determination of the latter is necessary follows from the results of~\cite{P_tz_2014,PhysRevA.108.042207}, in which it is argued that even in the presence of strong dependence of the measurement choices on hidden variables---as it is the case in  the setup of collider experiments--only an infinitesimal amount of randomness is necessary for genuine quantum non-locality to result in a violation of the Bell inequality. The initial conditions must then be fully specified for hidden variable models to overshadow the quantum effects and the vast amount of information required is to be confined within a finite volume of spacetime, thereby violating the Bekenstein bound~\cite{PhysRevD.23.287}. This fundamental principle, stating that initial conditions cannot be specified with arbitrary precision, is often overlooked in discussions  about determinism.\footnote{The burden for those who want to revive super-determinism would be to demonstrate that the number of digits required to set the initial conditions and enable predictability for a sufficiently long period (e.g., the age of the Universe), thereby also removing the however small amount of randomness required to violate appropriate Bell inequalities, is sufficiently small to fall below the Bekenstein bound. We also remark that even if one were to disregard the above argument and insist on entertaining such a philosophical stance, the contextuality requirement would still render full-blown super-determinism models rather unappealing.}}

\section{If they exist,  hidden variables will be new physics}

Hidden variables  were initially proposed to complement theories that were not fundamental, such as atomic physics.  On the other hand, particle physics delves into the most fundamental level of nature, where there is no room for hiding. This leads to speculations about whether the incompleteness of quantum mechanics can be rephrased as the incompleteness of the Standard Model interactions and particle content. While it is possible that a clever hidden variable model could reproduce the states identified by the tomographic procedure, regardless of the model, it will be contextual and non-local unless we consider the super-determinism choice, which we reject for the required infinite amount of information. If we attempt to construct a fundamental hidden variable model (an example of which, though non-relativistic, is Bohm's model~\cite{PhysRev.85.166}), we will face the challenge of doing so at the microscopic level and, therefore, by means of new particles or interactions that carry the information encoded in the hidden variables. These features cannot remain hidden in a fundamental theory and must manifest themselves as these new (currently elusive) particles or interactions with special properties. From the perspective of particle physics, the search for hidden variable models can then be framed as the search for physics beyond the Standard Model. 

As we contemplate the future of quantum tomography at colliders, two unexpected outcomes can emerge. If the experimental data indicated correlations surpassing the Tsirelson bound~\cite{Cirelson:1980ry}, the result would imply that non-locality surpasses the maximal amount that quantum mechanics can produce. All the tests conducted at colliders have so far confirmed quantum mechanics. Alternatively, the reconstructed state might not be an acceptable state within the Hilbert space of quantum mechanics. This could occur, \blue{for instance,} if the experimental values of certain cross sections violate unitarity. In such cases, the problem would be addressed by seeking the new physics effects responsible for the violation of unitarity rather than being taken as a signal of the failure of quantum mechanics. 

\section{Outlook}

\blue{In this paper, we have introduced a Bell non-locality study tailored to the spin correlations studied in collider experiments by means of quantum state tomography. The method allows for the reconstruction of the quantum state of the system under scrutiny and provides a witness of Bell non-locality. 
The collider results pertaining to Bell non-locality are inherently device-dependent owing to the experimental platform, which is characterized by a vulnerability reminiscent of the super-determinism loophole. As for the latter, we have stressed that super-determinism requires an absolute determination of the initial conditions incompatible with known fundamental limits in physics. 
%For this reason, the super-determinism loophole should be rejected.
}

\blue{Quantum state tomography offers new insights into the fundamental workings of quantum mechanics. Its methods have been successfully applied to analyze experimental data from colliders.  Specifically, analyses of $B$-meson decays at Belle II and LHCb~\cite{Fabbrichesi:2023idl}, Charmonium decays at BESIII~\cite{Fabbrichesi:2024rec,BESIII:2025vsr} and pions from kaon decays~\cite{Fabbrichesi:2025rqa} have been reinterpreted to achieve this. These processes have demonstrated \red{the possibility of witnessing} Bell non-locality with a significance exceeding $5\sigma$ \red{thus extending} the low-energy findings to encompass weak and strong interactions, as well as qutrits systems. Furthermore, the detection of entanglement in top-quark pair production data at the LHC has been achieved with a significance of $5\sigma$~\cite{ATLAS:2023fsd,CMS:2024pts,CMS:2024zkc}. } 
  
Analyses using $\tau$ lepton pairs~\cite{Privitera:1991nz,Ehataht:2023zzt}, top-quark pairs~\cite{Afik:2020onf,Fabbrichesi:2021npl,Severi:2021cnj,Aguilar-Saavedra:2022uye,Fabbrichesi:2022ovb,Dong:2023xiw,Han:2023fci} and vector gauge bosons~\cite{Barr:2021zcp,Aguilar-Saavedra:2022wam,Ashby-Pickering:2022umy,Fabbrichesi:2023cev,Morales:2023gow}  show that similar results can be expected in the upcoming years  for many other processes. The prospects of \red{studying} Bell \red{non-locality}  at future colliders has also been discussed~\cite{Altakach:2022ywa,Fabbrichesi:2023cev,Fabbrichesi:2024wcd,Morales:2023gow,Ma:2023yvd,Han:2025ewp}.

\vskip1cm
\noindent {\small
\textsc{Acknowledgements:} The authors would like to express their gratitude to Emidio Gabrielli for valuable discussions and two anonymous referees for their  contributions in enhancing the quality of the paper.  LM is supported by the Estonian Research Council under the RVTT3, TK202 and PRG1884 grants.}
\vskip2cm

\appendix
\section{Quantum tomography: a new take on loopholes}

Low-energy experiments with photons (or solid-state devices) can be considered more akin to proper Bell tests than those conducted at colliders because they can be formulated as black boxes.\footnote{The extent to which these experiments are device-independent must be evaluated on a case-by-case basis, see for example ~\cite{Rosset:2012yzz}.} In this black-box approach, the settings of the polarimeters serve as inputs, while the measured frequencies act as outputs (for a captivating illustration of this concept, refer to \cite{Mermin}). However, \textbf{local hidden variable models} pose a challenge to these tests by exploiting potential loopholes through which local variables could mimic Bell inequality violations. 

The applicability of these loopholes has been extended to encompass Bell non-locality measurements conducted at colliders. In these discussions, the  construction of suitable local hidden variable models is frequently done only in abstract terms.  The reason for this is straightforward: such models must simultaneously evade the stringent limits imposed by low-energy tests and bypass the high-energy tests. These requirements make the very definition of local hidden variable models suitable for collider experiments a daunting task.

Regardless of whether all loopholes are closed or not, establishing \blue{some form of} Bell non-locality in new settings, such as those provided by collider experiments, is of interest because, at the very least, it makes the construction of local hidden-variable models alternative to quantum mechanics more challenging. 
We believe there is more to it, though, due to the new features introduced by quantum tomography.

\red{Most of the} loopholes so far discussed in the literature exploit the incomplete experimental determination of the bipartite state that is provided in the traditional setup of a Bell test, notably in the two-photon system. Two directions, arbitrary but fixed, are defined for each of the two polarization measurements performed on each subsystem; correlations among these polarization measurements are then compared and the Bell inequality evaluated. These four directions are the context of the measurement; correlations are computed on some statistical distributions of the variables.

Loopholes bypass non-locality by encoding information on the measurement directions in additional variables. This way, the correlations entering the Bell inequality are altered once this information is transmitted. This is achieved either through an exchange of information that makes the direction $\vec n_i$, along which the first particle's spin is measured, known to the second particle whose spin is measured along the direction $\vec n_j$---as in the case of the space-time locality loophole---or by using this information to alter the events to be measured---as in the detection loophole. This key observation provides a new perspective on \blue{the applicability of} loopholes affecting Bell tests. It is crucial for the existence of loopholes that the measurement directions are set at some point, otherwise, no information can be used to manipulate correlations and consequently satisfy the inequality. This point is often overlooked in discussions of loopholes in Bell tests, especially when their effects are discussed in abstract terms.

Quantum tomography provides us with the state of the bipartite system under exam, a system that we could then probe to \blue{witness} the potential for Bell non-locality. If the state---which is non-local--- is given with sufficient precision, there is no way to bypass the mathematics that connects operators, states, and expectation values. The presence of entanglement and  \red{Bell non-locality} can then be straightforwardly established through the mentioned observables, within the uncertainties proper of the tomographic measurements. There is no room for loopholes to act: loopholes do not affect states, quantum tomography nor quantum mechanics.  \red{Most} loopholes  exploit the fact that only partial information is made available in a Bell test where the freedom of manipulating the setting necessarily leaves room for the loopholes to slip in.   

\blue{Since in the literature, there exist various attempts to extend the discussion of these loophole in the framework of collider physics,} in the following we briefly review the different loopholes affecting Bell tests one at a time, showing how the tomographic procedure made available by collider experiments allows to disregard them. 

\subsection{Space-time locality}
\begin{quote}
This loophole exploits the space-time arrangement of the events involved in a Bell test. Communication pertaining to the direction settings used in the  polarization measurements performed on the two subsystems is possible in a local manner if these settings are fixed before the source emits the entangled particles, or if one setting and measurement is performed in the past cone of the second one. In both  cases, the Bell inequality is violated because of the local exchange of information~\cite{Bell:1964}.
\end{quote}

Local hidden variables theories can bypass the non-separability by using interactions to transmit information, provided the measurements are taken at time-like intervals and the hidden variables carry information about the direction setting of one of the two photon measurements. In the traditional Bell test setting, this loophole has been closed by first arranging the polarization measurement to be space-like separated and by letting the polarization directions be determined by a quantum random number generator~\cite{Aspect:1982fx,Weihs:1998gy,Tittel:1998ja}.

At collider experiments, the condition of space-like separation between the polarization measurements requires, at first blush, particles with the same lifetime and mass. Yet this restriction is actually not necessary because the time-like separation between production and measurement points is only an issue for  experiments with the traditional settings. At colliders, in fact, no specific setting is decided beforehand as the whole density matrix is fully reconstructed. The directions involved in \red{witnessing Bell non-locality} are decided only afterwards, when the reconstructed state is pulled through the machinery of quantum mechanics to compute the required correlation values. There is no way for the state at the production point, nor for the two subsystem composing the final state, to know this choice in advance. For this reason, even for two decays taking place one inside the future cone of the other---as in the case of two particles with very different masses and decay lengths---the loophole cannot be invoked.

\subsection{Detection}
\begin{quote}
It is possible that the correlations, as detected in the experiment, are only due to the particular subset of recorded events, thereby violating the assumption of fair sampling. Though the selected events show a violation of a Bell inequality, if all events were detected the Bell inequality would actually be respected. If some events are not recorded, extra correlations could be hiding there~\cite{Weihs:1998gy}. In~\cite{PhysRevD.35.3831} it was shown that the efficiency for the detection of the photons should be at least 83\%.
\end{quote}

The loophole has been closed in low-energy tests in~\cite{Rowe:2001kop,Matsukevich:2008zz,Ansmann:2009ims}. This and the locality loopholes have been closed simultaneously in~\cite{Hensen:2015ccp,Li:2018smx,Stevens:2015awv,Giustina:2015yza}.

This problem is usually 
 addressed in collider experiments by the requirement of fair sampling.\footnote{Fair sampling is necessary in high-energy physics also in ordinary cross section measurements---otherwise you might think that, for instance, supersymmetry is hiding in the many events that have not been detected.} This assumption, which is precisely the one violated by the loophole,  is not necessary because examining the conditions under which the loophole might work (for instance in~\cite{Gisin:1999nr}) reveals that this loophole is also closed by the collider settings, without resorting to the fair sampling assumption. The missing events, the basis of the loophole, are not generic ones because  in this case they would only reduce the significance of the test results. Most events that do not violate the Bell inequality must be excluded, while those that do violate it must be retained. The selection of events is conducted by means of the directions chosen in the experimental settings. However, since no  directions are chosen during the tomographic reconstruction of the quantum state, the loophole cannot even be formulated.

\vskip3cm
\subsection{Freedom of choice}
\begin{quote}
In this loophole, an extension of the locality one, the hidden-variable distribution $\eta$ depends on the directions used in the Bell test for the polarization measurements. When calculating
\be
 {\mathcal P} (\uparrow_{\hat n_i}; \uparrow_{\hat n_j})=  \int  \di \lambda \, p_\lambda (\uparrow_{\hat n_i}; \uparrow_{\hat n_j}) \eta (\lambda| \uparrow_{\hat n_i}; \uparrow_{\hat n_j})\, ,
\ee
one should take into account possible correlations between the hidden variables and the detector settings~\cite{Brans:1987}, which would modify the above relation thereby allowing a hidden variable theory to pass the Bell test.
This loophole essentially proposes a limited form of a super-deterministic model in which, however, only a tiny amount of information needs to be exchanged~\cite{Friedman:2018byq}.
\end{quote}

This loophole is considered impossible to close through scientific methods in the traditional setup (although see the objection presented in the \red{main text}). In low-energy tests it has been made more and more unlikely by making the choice of detector settings be determined by a large ensemble of pseudo-random events (canvassed from a large sampling of human choices)~\cite{BIGBellTest:2018ebd} or events that are in the very (cosmological) far past~\cite{Rauch:2018rvx}. 

At collider experiments, no detector settings are chosen. There is therefore no information that the hidden variables can carry to alter the reconstruction of the state which proceeds always in the same manner. Likewise, the loophole cannot alter the subsequent probing of correlations and \red{the witnessing of Bell non-locality}, which are solely regulated by quantum mechanics.

\subsection{Memory} 
\begin{quote}
If the measurements are repeatedly made, a local hidden variable theory could exploit the memory of past measurement settings and outcomes to increase the correlations and yield a violation of the Bell inequality in the upcoming tests~\cite{PhysRevA.66.042111}.
\end{quote}
The loophole was identified in~\cite{PhysRevA.66.042111} where it was also shown how the loophole is less and less effective as the number of measurements grows.

At collider experiments, the measurements are intrinsically random because the events originate from particles decaying at different times and positions in the beam-pipe or the detector. Moreover, the memory that is passed on is about the directions of the measurements and, again, these are not set by the tomographic procedure. Finally, the number of events collected at colliders is large enough for the loophole to be closed.

\subsection{Coincidence} 
\begin{quote}
In many experiments, especially those based on photon polarizations, pairs of events in the two sides of the experiment are only identified as belonging to a single pair after the experiment is performed, by judging whether or not their detection times are close enough one to another. This generates a new possibility for a local hidden variables theory to fake quantum correlations: altering the detection time of each of the two particles according to some relationship between hidden variables carried by the particles and the detector settings encountered at the measurement station~\cite{Larsson:2003efo}.
\end{quote}
The loophole was identified and closed at low-energy in the same work~\cite{Larsson:2003efo}.

Coincidence is  well verified in collider experiments by the kinematic reconstruction performed event by event. Moreover, the amount of misidentifications is routinely included in the uncertainty of the measurements.

%\newpage
\begin{multicols}{2}
%\twocolumn  
\small
\bibliographystyle{JHEP}   
\bibliography{bell} 

\end{multicols}

\end{document}